\documentclass{article}
\usepackage[utf8]{inputenc}
\usepackage{geometry}
\usepackage[table]{xcolor}
\usepackage{hyperref}
\usepackage{multirow}
\usepackage{graphicx}
\usepackage{booktabs,subcaption, longtable, array, float}
\usepackage{anyfontsize}
\usepackage{dirtytalk}
\usepackage{amsmath,amssymb}
\usepackage{lscape}
\usepackage{adjustbox}
\usepackage{authblk}
\usepackage[toc]{appendix}

\graphicspath{{Figures/}}

\usepackage[sorting=none, backend=biber]{biblatex}
\title{The social stratification of internal migration and daily mobility during the COVID-19 pandemic}

\author[1,*]{Erick Elejalde}
\author[2,3,4,*]{Leo Ferres}
\author[2,3]{Víctor Navarro}
\author[2,3]{Loreto Bravo}
\author[5]{Emilio Zagheni}
\affil[1]{L3S Research Center, Leibniz University Hannover, Germany}
\affil[2]{Faculty of Engineering, Universidad del Desarrollo, Santiago, Chile}
\affil[3]{Telefónica R\&D, Santiago, Chile}
\affil[4]{ISI Foundation, Turin, Italy}
\affil[5]{Max Planck Institute for Demographic Research, Rostock, Germany}
\affil[*]{Corresponding authors: E-mail:\href{mailto:elejalde@l3s.de}{elejalde@l3s.de}, \href{mailto:lferres@udd.cl}{lferres@udd.cl}}

\date{}

\begin{document}
\setlength{\extrarowheight}{1em}
\maketitle
\begin{abstract}
This study leverages mobile phone data for 5.4 million users to unveil the complex dynamics of internal migration and daily mobility in Santiago de Chile during the global COVID-19 pandemic, with a focus on socioeconomic differentials. Major findings include an increase in daily mobility among lower-income brackets compared to higher ones in 2020. In contrast,  long-term relocation patterns rose primarily among higher-income groups. These shifts indicate a nuanced response to the pandemic across socioeconomic strata. Unlike in 2017, economic factors in 2020 influenced a change not only in the decision to emigrate but also in the selection of destinations, suggesting a profound transformation in mobility behaviors. Contrary to expectations, there was no evidence supporting a preference for rural over urban destinations despite the surge in emigration from Santiago during the pandemic. The study enhances our understanding of how varying socioeconomic conditions intersect with mobility decisions during crises and provides valuable insights for policymakers aiming to enact fair, informed measures in rapidly changing circumstances.
\end{abstract}

\section{Introduction}
\label{sec:intro}
Migration, a universal and multifaceted phenomenon, encompasses the movement of entities --ranging from people and animals to data and ideas-- across various locations. This movement manifests in forms as diverse as long-term relocations and daily mobility, each constituting critical aspects of the broader migratory landscape. Human migration is particularly driven by a myriad of factors, such as the quest for better living conditions, life-course transitions, the urgency to flee conflict, or environmental variables like pandemics \cite{IOM_22-rp, green2017understanding, Bernard_2016}. Within this scope, external or international migration involves relocating across national boundaries, motivated by circumstances like economic opportunities, political unrest, and environmental conditions. In contrast, internal migration occurs within a single country, usually defined by administrative political divisions like states, provinces, or districts \cite{bell2002cross}. Additionally, daily mobility represents the routine, short-range movement primarily confined within smaller spatial areas like cities or districts \cite{vinceti2020lockdown}. Both long-term internal migrations and daily mobility patterns are integral to understanding the spatio-temporal transformations in social systems and human settlements, particularly during periods of rapid change such as pandemics \cite{rees2017impact, gozzi2021estimating}.

Internal migration overwhelmingly dominates global movement, accounting for the majority of relocations \cite{skeldon2017international}. This prevalence underscores the urgency of effective monitoring to understand its multifaceted dynamics, including the spatio-temporal transformations it brings to social systems and human settlements through mechanisms like population concentration and deconcentration \cite{rees2017impact}. Alongside long-term migrations, daily mobility patterns also exert a transformative influence, especially during crises such as pandemics \cite{gozzi2021estimating}. Importantly, these mobility behaviors are not universally experienced; they vary across different societal groups and are significantly influenced by policies and other factors \cite{gauvin2020gender, duenas2021changes}. Consequently, advancing methods to monitor both long-term and daily internal migrations is pivotal for policymakers to enact equitable and timely interventions, particularly when confronting rapidly evolving situations \cite{skeldon2017international}.

The utilization of mobile phone data has instigated a seismic shift in the study of human mobility, supplanting traditional methods that primarily relied on population censuses and suffered from a paucity of high-frequency, detailed data \cite{skeldon2017international}. This innovative data source provides unprecedented, real-time insights into both long-term internal migration and daily mobility patterns, thereby heralding a transformative era in our understanding of human movement \cite{rees2017impact}. The granular nature of mobile phone data allows for enhanced temporal and spatial resolution, facilitating the tracking, analysis, and visualization of migration flows like never before. This not only enables the observation of mobility in ``near real-time'' but also aids in capturing sudden shifts triggered by factors such as political upheaval or pandemics. As a result, mobile phone data not only fills the gaps left by traditional methods but also offers the potential to standardize cross-national comparisons by homogenizing the types of data collected, measurement intervals, and spatial frameworks \cite{bell2002cross, Bell_2015}.

Research on human mobility during the COVID-19 pandemic has mostly centered on daily movements, such as the effectiveness of lockdown measures \cite{vinceti2020lockdown} or the relationship between mobility and pandemic progression \cite{gozzi2021estimating, mena2021socioeconomic}. This paper extends this focus to encompass long-term internal migration during the pandemic. In particular, we explore the role of socioeconomic factors in both daily and long-term internal migration, using the comunas of Santiago de Chile as a case study. We examine the transformation in mobility dynamics across different economic strata from 2017 to 2020, showcasing that daily mobility has undergone a significant shift, especially among lower-income brackets. This highlights the need for a more nuanced understanding of mobility patterns during crises, informed by both long-term migration and daily movement metrics \cite{rees2017impact, bell2002cross}.

In particular, we explore questions related to internal migration patterns and their connection to socioeconomic factors during the global COVID-19 pandemic. We delve into the dynamics of long-term mobility during this period, focusing on the role of varying socioeconomic strata \cite{Gaag_2008}. Alongside long-term mobility, we also scrutinize daily mobility patterns to draw comparative insights between the two, particularly in relation to socioeconomic variables. Our investigation extends to the changing relationship between socioeconomic levels and migration patterns, capturing the shift in correlation between the average socioeconomic level and the percentage of the population migrating from and to each comuna of Santiago de Chile in 2017, 2020, and 2022. In the context of Chile, where a ``comuna'' represents the smallest administrative unit (similar to a ``municipality'' in other countries), we specifically spotlight migration behaviors in high-income comunas, which have seen a notable uptick in emigration coupled with low variability in destination selection. Concurrently, we assess individual preferences regarding the population density of chosen destinations, with a keen focus on tendencies towards rural locales. Our study, thus, aims to offer a nuanced understanding of the interplay between socioeconomic and environmental factors influencing both long-term and daily mobility during global crises, leveraging the granular capabilities of mobile phone data.

\section{Materials and Methods}
\label{sec:methods}
Chile is widely acknowledged as a heavily centralized country from various perspectives. A 2013 study revealed that, in proportion to its size, population, and economic development, Chile was the most centralized country in Latin America \cite{vonBaer2013descentralizado}. Data obtained from the National Institute of Statistics (INE) indicates a total estimated population of 19.4 million for the entire country in 2020, with 8.1 million (41\%) concentrated in the Metropolitan Region, where the capital, Santiago (SCL), is located \cite{ine_cl_proyecciones_anuales}. Notably, this region is among the smallest of the 16 regions in Chile, resulting in a densely populated area. 

This makes Santiago comparable, in raw numbers, to places like Hong Kong, Baghdad, or even New York City. 
With 32 comunas (the smallest political division in Chile) and a diverse range of socioeconomic conditions, SCL presents an interesting case study for internal migration dynamics during a crisis period. SCL has over 95\% of its area urbanized and contributed over half of the confirmed COVID-19 cases and over 60\% of the deaths in the first six months of the pandemic \cite{mena2021socioeconomic}. 

As COVID-19 swept across the globe in the early months of 2020, there were emerging reports of an 'urban exodus' phenomenon trending not only in Chile \cite{urban2rural_news_chile} but several other big cities \cite{urban2rural_news_USA, urban2rural_news_canada}. During the first waves of infections, when there was still not much information about the virus, densely interconnected cities bore the brunt of their impact. Some newspaper headlines questioned the future of urban areas, even beyond the end of the pandemic \cite{weforum}. We set to investigate the dynamics of intracity daily mobility, and long-term long-distance relocation for a large metropolitan area during COVID-19. Our approach is to use a comparative analysis to identify changes in mobility in 2020 compared to 2017 for a ``during pandemic'' period and the same in 2022 for the ``recovery period''. We study the mobility metrics in relation to socioeconomic status indicators for the comunas of origin and destination.

\subsection{Long-term relocation}
Previous analyses on long-term relocation during COVID-19 are based on social media data \cite{rowe2022urban} or official records from authorities \cite{gonzalez2022understanding, vogiazides2022internal, fielding2021covid}. This represents a limitation for its applicability to regions with lower social media penetration or slow/expensive cost of polling official values.

We propose a model for long-term relocation based on the analysis of anonymized eXtended Detail Records (XDRs). 
For each studied year (2017, 2020, and 2022), we collected eight months of XDRs for approximately 1.3 million devices per year in the Metropolitan Region. This data captures interactions, such as packet requests, between devices and antennas. A data entry in our dataset can be represented as a tuple $<d,t,
a>$, denoting a packet request from device $d$ to antenna $a$ at time $t$. We estimate the location of device $d$ based on the fixed latitude and longitude of the antenna it connects to.

We used the second week of March as the baseline for the initial location because this is the week before the return to school in Chile. So, most people should be back from vacation at their primary residence.
We then estimate the home location for each device using the home detection algorithm proposed in a previous work \cite{pappalardo2021evaluation}. This algorithm looks for each device's most used cell tower during night-time on weekdays. We assign antennas to correspondent comunas according to their position.
Finally, for each device, we calculated its home comuna per week from Mar $1^{st}$ until Nov $30^{th}$ of each year. If the statistical mode of home in the four weeks of November was in a different region than their March home, we assumed that the device had migrated.
To gauge the level of migration activity, we count individuals relocating to and from each comuna. Also, to assess the geographical impact of migration, we quantified the net migration rate, representing the overall balance between incoming and outgoing migration flows while accounting for population size \cite{Lieberson_1980}.

Given our main interest in urban mobility dynamics, we restrict our analysis to the capital city Santiago (SCL), formed of 32 comunas 95\% urbanized and with a population of around 6 million people \cite{ine_cl_proyecciones_anuales}. For SCL, our dataset has 979.1K, 989.6K, and 955.8K devices for the years 2017, 2020, and 2022 respectively. For the immigration analysis, we collect data on over 2 million devices each year originally located outside the Metropolitan Region. In order to protect the privacy of device owners, we exclusively examine and present results that are both anonymized and aggregated. Moreover, we did not utilize or have access to any additional user information, such as gender or age.

\begin{figure*}[ht!]
    \centering
    \begin{subfigure}[t]{0.32\textwidth}
        \centering
        \includegraphics[width=\linewidth]{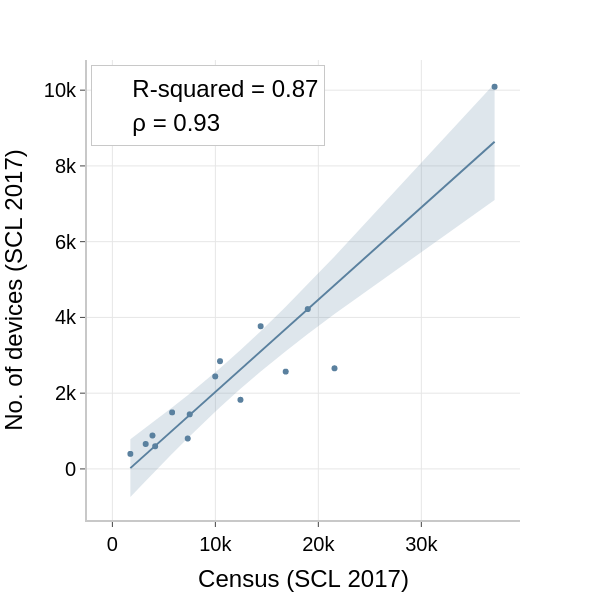} 
        \caption{Immigration - Regions of Origin} \label{fig:imm_region_val_2017}
    \end{subfigure}
    \hfill
    \begin{subfigure}[t]{0.32\textwidth}
        \centering
        \includegraphics[width=\linewidth]{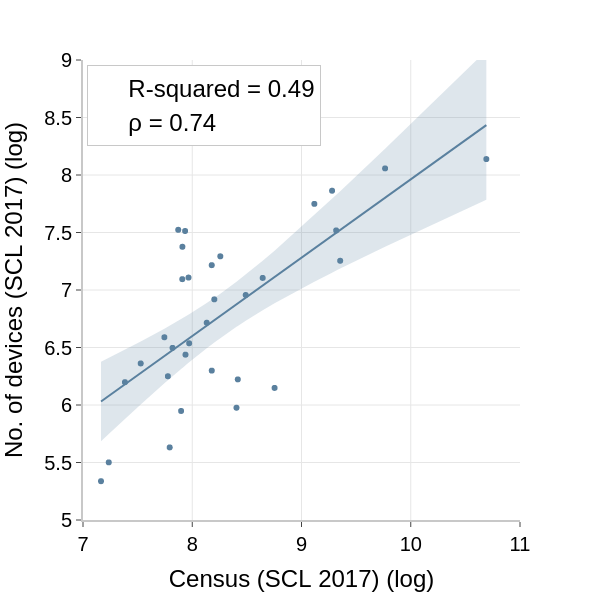} 
        \caption{Immigration - Comunas of Destination} \label{fig:imm_MR_val_2017}
    \end{subfigure}
    \hfill
    \begin{subfigure}[t]{0.32\textwidth}
        \centering
        \includegraphics[width=\linewidth]{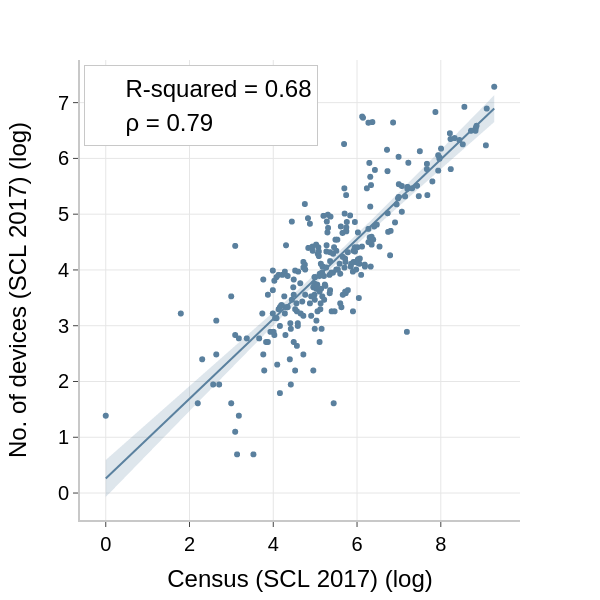} 
        \caption{Immigration - Comunas of Origin} \label{fig:imm_comuna_val_2017}
    \end{subfigure}
    \begin{subfigure}[t]{0.32\textwidth}
        \centering
        \includegraphics[width=\linewidth]{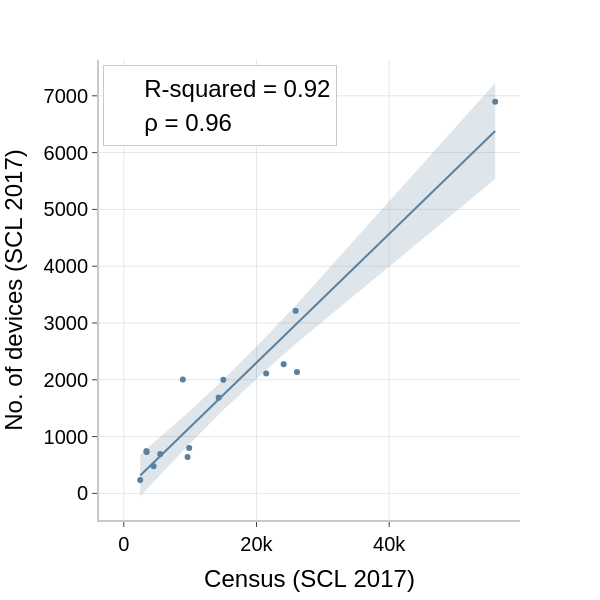} 
        \caption{Emigration - Regions of Destination} \label{fig:em_region_val_2017}
    \end{subfigure}
    \hfill
    \begin{subfigure}[t]{0.32\textwidth}
        \centering
        \includegraphics[width=\linewidth]{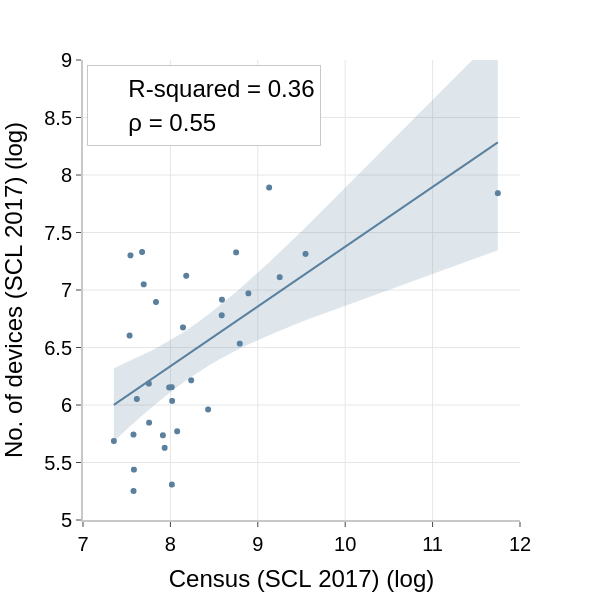} 
        \caption{Emigration - Comunas of Origin} \label{fig:em_MR_val_2017}
    \end{subfigure}
    \hfill
    \begin{subfigure}[t]{0.32\textwidth}
        \centering
        \includegraphics[width=\linewidth]{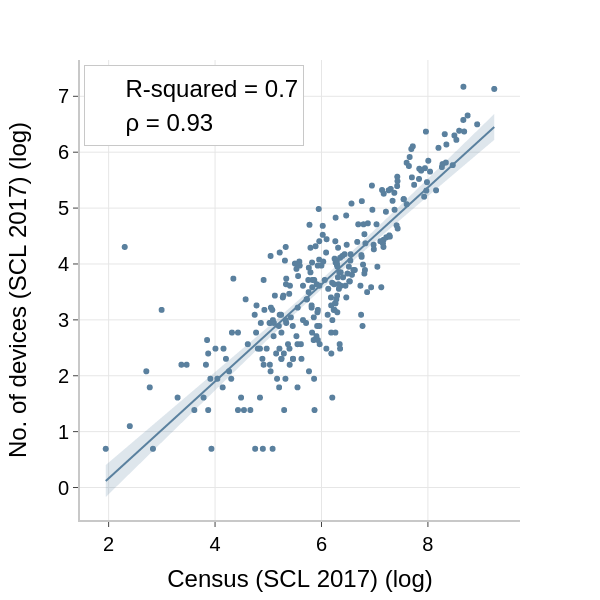} 
        \caption{Emigration - Comunas of Destination} \label{fig:em_comuna_val_2017}
    \end{subfigure}
    \caption{Comparing the movement based on the Internal Migration Mobile Model (Mar.-Nov., 2017) and the movement reported by the National Census 2017. \ref{fig:imm_region_val_2017} and \ref{fig:em_region_val_2017} represent movement between other regions and the Metropolitan Region (MR). \ref{fig:imm_MR_val_2017} and \ref{fig:em_MR_val_2017} represent movement into and from comunas in the MR. \ref{fig:imm_comuna_val_2017} and \ref{fig:em_comuna_val_2017} represent the migration between the MR and other comunas outside this region.}
    \label{fig:censo_mobile_2017} 
\end{figure*}

We further validate that our definition can be used to approximate the movement from SCL to other regions in Chile. For this, we compared our measurements for 2017 against the migration information from the National Census 2017. Table~\ref{tab:census_model_corr_2017} summarizes the correlation between internal migration values from the census and the Internal Migration Mobile Model at various granularity levels. At the regional level (i.e., in- and outflow between other regions and the Santiago de Chile (SCL) - Regions $\leftrightarrows$ SCL), we get a very high correlation ($.93$ and $.96$). We also analyzed the flow between other comunas and SCL as a whole (comunas $\leftrightarrows$ SCL), and between the comunas in the MR and the rest of the country as a whole (Country $\leftrightarrows$ SCL-comunas). Even for the finest granularity offered by the Census (comunas $\leftrightarrows$ SCL-comunas), our model shows relatively high correlations ($.63$ and $.48$).  Additional details can be found in the SI Appendix (Figure \ref{fig:censo_mobile_2017})

\begin{table*}
\centering
\caption{Correlation between internal migration values from the National Census 2017 and the Internal Migration Mobile Model (Mar.-Nov., 2017). In both cases, we only consider movements between locations inside Santiago de Chile (SCL) and locations outside SCL (e.g., not considering migration inside the SCL).}
\label{tab:census_model_corr_2017}
\begin{tabular}{lcc}
Correlation & Immigration & Emigration \\
\midrule
Regions $\leftrightarrows$ SCL & $r(13)=.93, p< .001$ & $r(13)=.96, p< .001$ \\
comunas $\leftrightarrows$ SCL & $r(280)=.79, p< .001$ & $r(280)=.93, p< .001$ \\
Country $\leftrightarrows$ SCL-comunas & $r(30)=.74, p< .001$ & $r(30)=.55, p< .001$ \\
comunas $\leftrightarrows$ SCL-comunas & $r(5948)=.64, p< .001$ & $r(5161)=.51, p< .001$ \\
\bottomrule
\end{tabular}
\end{table*}

\subsection{Daily Mobility}
Alongside long-term migrations, daily mobility patterns also exert a transformative social influence, especially during crises such as pandemics. For this short-term mobility analysis, we use a dataset of XDRs collected for the same population (i.e., comunas in SCL) in 2020, comprising 5.4 million users country-wise \cite{Pappalardo2023}. From the XDRs, the authors produce three epidemiologically relevant metrics: the Index of Internal Mobility, which quantifies the amount of mobility within each comuna of the country; the Index of External Mobility, quantifying the mobility between comunas; and the Index of Mobility (IM), which considers any movement, both within and between comunas.
The data used to calculate the daily mobility index, as well as the active quarantine periods, are available for download from the general repository of the Ministry of Science of Chile \cite{im_2020_data, quarantine_2020_data}.

Similar to long-term relocation, we are interested in studying the changes in the mobility dynamics during a period of crisis. We analyze the reduction in mobility for all three indices during the year (March to November) compared to the same baseline, i.e., the second week of March. Additionally, we repeat the analyses considering only periods of quarantine implemented at the comuna level and only periods of no-quarantine.

\subsection{Socioeconomic aspects}
Our primary source for the socioeconomic metrics is the Socioeconomic Characterization Survey (CASEN) \cite{casen2020} - the main household survey in Chile. The CASEN survey is designed to be representative at the national level, by urban and rural geographic areas, and also at the regional level. However, it is worth noting that CEPAL (United Nations Economic Commission for Latin America and the Caribbean) is tasked with making corrections for non-responses, addressing missing income data, and rectifying potential underreporting or overreporting of various income categories prior to making the databases officially accessible to the public \cite{cepal_sae}. They use statistical techniques and probabilistic models to produce disaggregations for groups of interest, known as small area estimation (SAE) techniques \cite{rao2015small}.
For the synthetic model of the SAE estimates carried out by the Ministry of Social Development and Family, besides the results from the survey, there is information from administrative records collected by the public sector at the commune level and information from the Population Census \cite{molina2019desagregacion}. From the CASEN, we obtain estimates for the average income decile and the percentage of poverty per comuna \cite{pobreza_comunas}.

For people who decided to migrate between urban areas, we are interested in the differences in quality of life between origin and destination. Following the hypothesis of socioeconomic components influencing the migration dynamics during the pandemic, we expect the migrant to try to move laterally in terms of urban amenities to maintain a similar quality of life. 
Here, we use the Index of Quality of Life in Urban Areas (ICVU) \cite{icvu_2020}. ICVU is a synthetic index employed to assess and compare the relative quality of urban life in Chilean communes and cities. This index relies on a collection of variables that pertain to six dimensions, reflecting the status of public and private goods and services provided to the resident population, as well as their socio-territorial consequences. This evaluation spans from larger cities to intermediate urban centers (more than 50K inhabitants) and encompasses the metropolitan scale.
Given that the index is limited to urban areas, it only covers 99 comunas in Chile (including the 32 comunas in SCL).

For each pair of origin-destination included in the ICVU, we calculate the score difference. Following, we compute for average ICVU difference for each comuna of origin weighted by the percentage of migrants that moved between each pair of comunas. Finally, we regress these average ICVU differences to analyze how different origins deviate from the trend, especially in connection to their economic status indicators.

\subsection{Urban-rural mobility}
Although the trend of deurbanization in major cities did not start with the COVID-19 pandemic, health concerns and stricter mobility restrictions in 2020 reportedly accelerated the process \cite{urban2rural_news_argentina}.

For a more quantitative insight, we calculated the average rurality of migration destinations for each SCL comuna, weighted by the percentage of emigrants. Specifically, the difference, $\Delta R_x$, between 2017 and the years 2020 and 2022 is computed as:

\[
   \Delta R_x = R^T_x - R^{T_0}_x 
\]
\[
   R^T_x = \frac{1}{M^T(x)}\sum_{y \in D} m^T(x,y)r(y)
\]

\noindent where:
\begin{itemize}
    \item $D$: Set of destination comunas.
    \item $M_T(x)$: Total migration from comuna $x$ to $D$ in year $T$.
    \item $m^T(x,y)$: Migration from comuna $x$ to $y$ in year $T$.
    \item $r(y)$: Rurality percentage of comuna $y$.
    \item Years $T$ include 2020 and 2022, with $T_0$ being 2017.
\end{itemize}

To estimate the percentage of rurality for each comuna ($r(y)$), we use the data from the CASEN \cite{casen2020}. Although it is not designed to be representative at the comuna level, it gives us a good approximation of the rural composition of these areas \cite{rurality_chile}. Here, we use the percentage of households annotated as rural for each comuna.

Nevertheless, rurality is a changing concept and can be operationalized in different ways, which may bias comparisons across countries \cite{nelson2021definitions}. Thus, for a finer-grain analysis of destination preference in terms of urbanization, we further investigate the emigration from each comuna in SCL in relation to the average of their destinations' population density \cite{ine_cl_proyecciones_anuales, rurality_chile}. For the average population density, we again weigh each destination by the percentage of the emigration from the corresponding comuna of origin. The analysis in terms of population density complements the rurality index above as it might give us some nuances inside each class (i.e., urban and rural). For example, one of the main destinations for emigration from Santiago is the region of Valparaiso, which is also highly urbanized. In this case, using the population density to compare multiple urban destinations can uncover additional insights into destination preferences or affordability.

Furthermore, rural and distant regions might be deficient in the necessary infrastructure and amenities required to accommodate new arrivals from urban areas \cite{weeden2022urban}. For a preliminary analysis of the potential impact of the increased emigration from Santiago, we calculate the percentage that the estimated emigration represents for the hosting populations. A sudden increase of several percentage points in the population of a rural community may represent a challenge, and more so during a period of crisis \cite{peters2020community}.

\section{Results}  

\begin{figure*}[ht!]
    \centering
    \begin{subfigure}[t]{0.3025\textwidth}
        \centering
        \includegraphics[width=\linewidth]{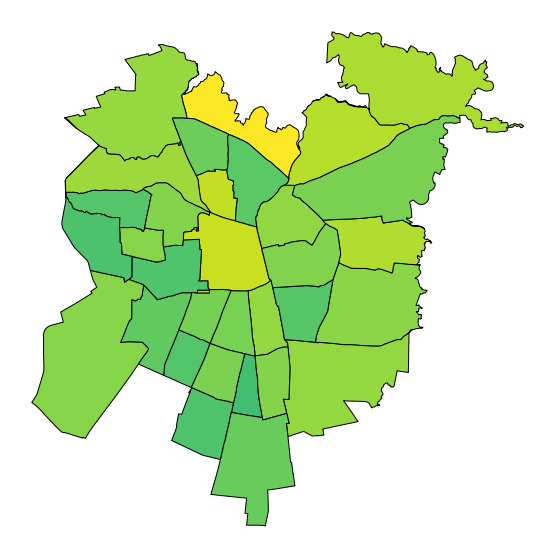} 
        \caption{Net migration rate SCL 2017} \label{fig:net_migration_map_2017}
    \end{subfigure}
    \hfill
    \begin{subfigure}[t]{0.3025\textwidth}
        \centering
        \includegraphics[width=\linewidth]{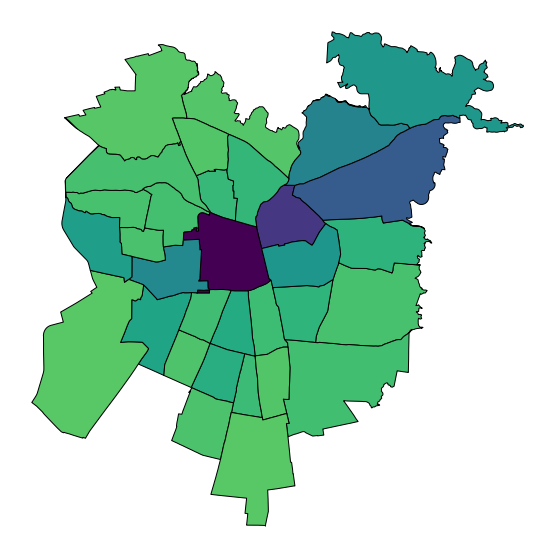} 
        \caption{Net migration rate SCL 2020} \label{fig:net_migration_map_2020}
    \end{subfigure}
    \hfill
    \begin{subfigure}[t]{0.355\textwidth}
        \centering
        \includegraphics[width=\linewidth]{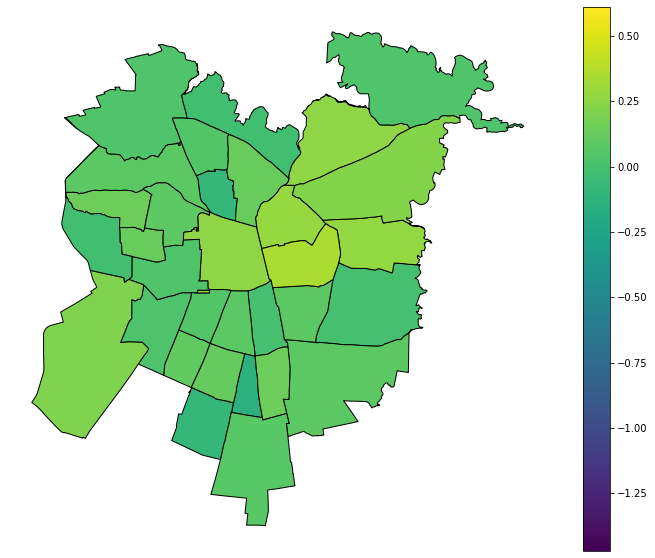} 
        \caption{Net migration rate SCL 2022} \label{fig:net_migration_map_2022}
    \end{subfigure}
    \begin{subfigure}[t]{0.32\textwidth}
        \centering
        \includegraphics[width=\linewidth]{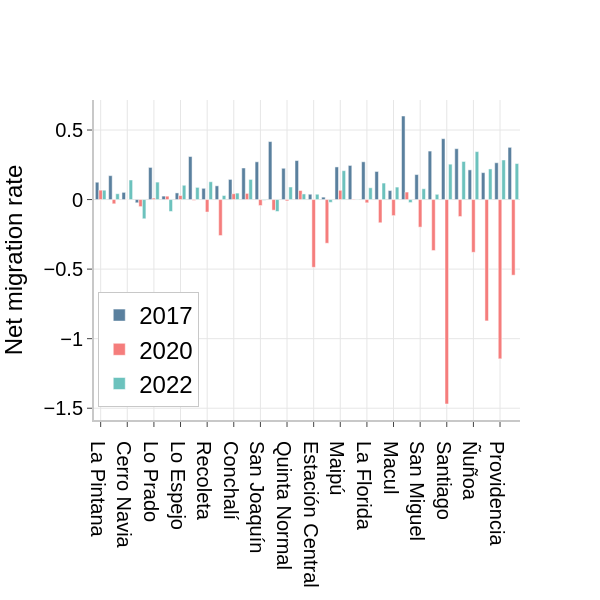} 
        \caption{Net migration rate per year} \label{fig:net_migration_per_year}
    \end{subfigure}
    \hfill
    \begin{subfigure}[t]{0.32\textwidth}
        \centering
        \includegraphics[width=\linewidth]{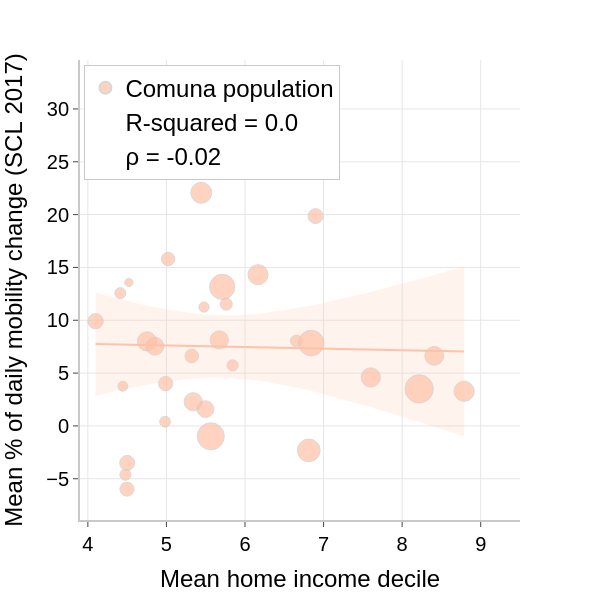} 
        \caption{Daily Mobility Index change in SCL (2017)} \label{fig:daily_IM_change_2017}
    \end{subfigure}
    \hfill
    \begin{subfigure}[t]{0.32\textwidth}
        \centering
        \includegraphics[width=\linewidth]{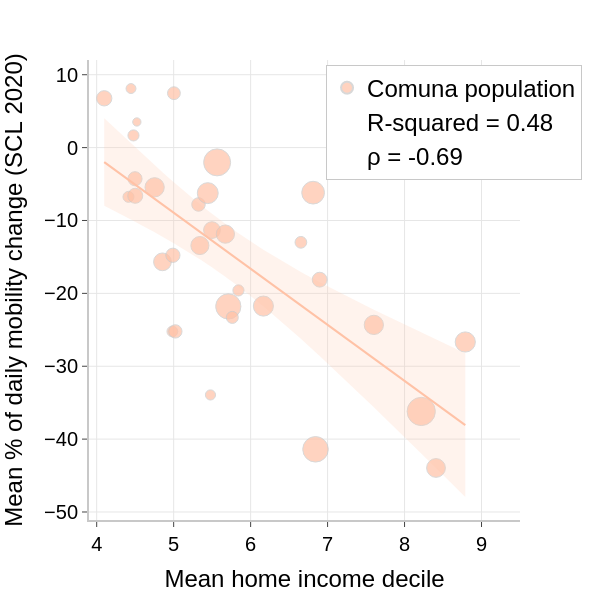} 
        \caption{Daily Mobility Index change in SCL (2020)} \label{fig:daily_IM_change_2020}
    \end{subfigure}
    \begin{subfigure}[t]{0.32\textwidth}
        \centering
        \includegraphics[width=\linewidth]{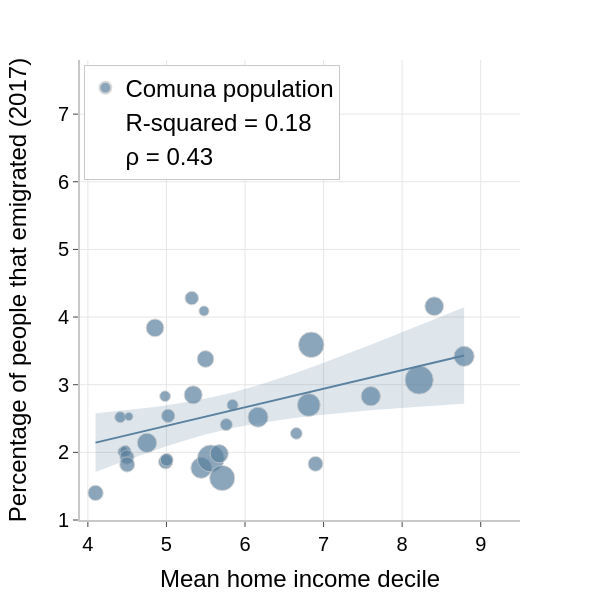} 
        \caption{Percentage of emigration from SCL vs. average home income decile (2017)} \label{fig:em_MR_com_2017_income}
    \end{subfigure}
    \hfill
    \begin{subfigure}[t]{0.32\textwidth}
        \centering
        \includegraphics[width=\linewidth]{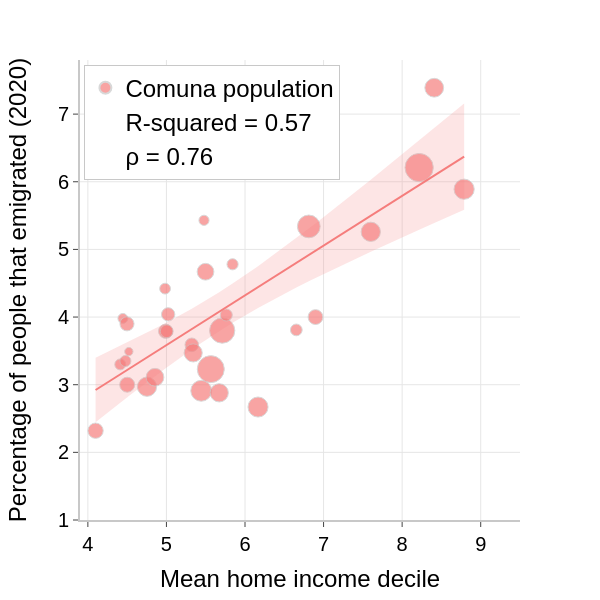} 
        \caption{Percentage of emigration from SCL vs. average home income decile (2020)} \label{fig:em_MR_com_2020_income}
    \end{subfigure}
    \hfill
    \begin{subfigure}[t]{0.32\textwidth}
        \centering
        \includegraphics[width=\linewidth]{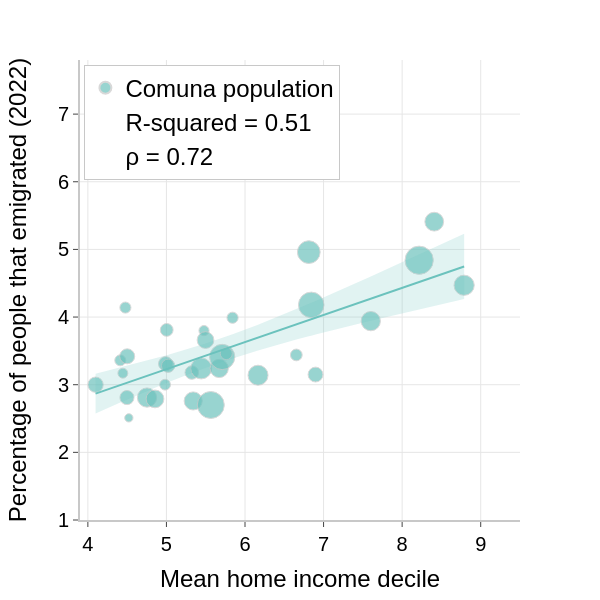} 
        \caption{Percentage of emigration from SCL vs. average home income decile (2022).} \label{fig:em_MR_com_2022_income}
    \end{subfigure}
    \caption{(\textbf{Origin}): Analysis of the emigration from the Metropolitan Region (SCL) before, during, and after the COVID-19 pandemic. Comunas from SCL in the X axis are sorted (from left to right) by ascending comunas' average household income decile.}
    \label{fig:emigration_origin} 
\end{figure*}

\begin{figure*}[ht]
    \centering
    \begin{subfigure}[t]{0.24\textwidth}
        \centering
        \includegraphics[width=\linewidth]{figures/daily_IM_change.png} 
        \caption{IM change 2020} \label{fig:IM_re}
    \end{subfigure}
    \hfill
    \begin{subfigure}[t]{0.24\textwidth}
        \centering
        \includegraphics[width=\linewidth]{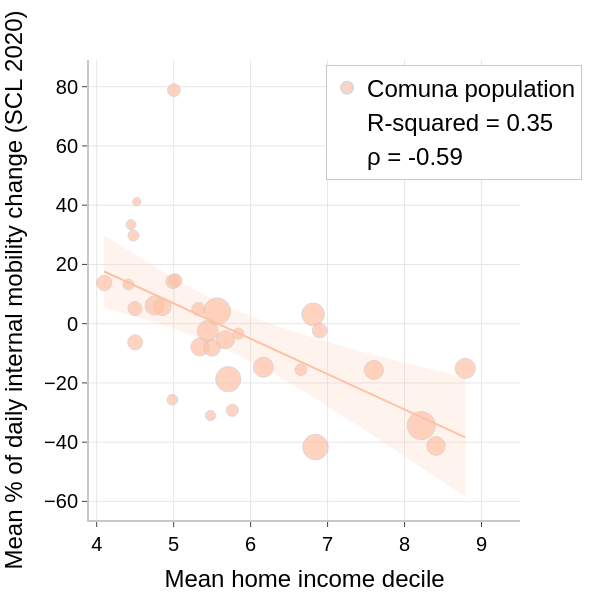} 
        \caption{IM internal change 2020} \label{fig:IM_internal_change}
    \end{subfigure}
    \hfill
    \begin{subfigure}[t]{0.24\textwidth}
        \centering
        \includegraphics[width=\linewidth]{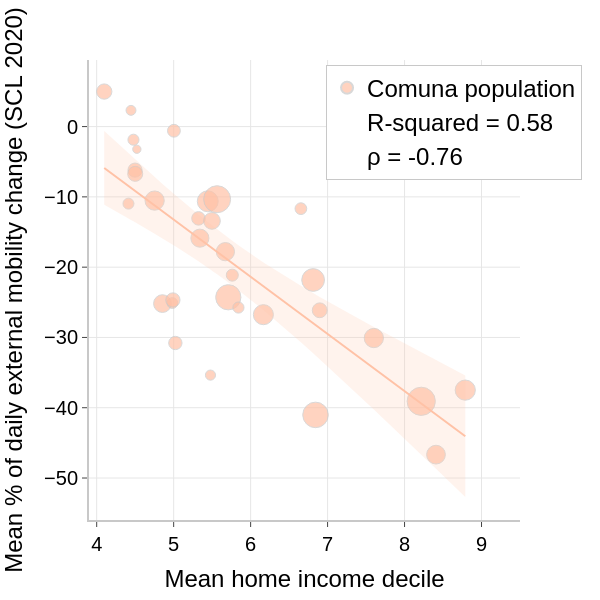} 
        \caption{IM external change 2020} \label{fig:IM_external_change}
    \end{subfigure}
    \hfill
    \begin{subfigure}[t]{0.24\textwidth}
        \centering
        \includegraphics[width=\linewidth]{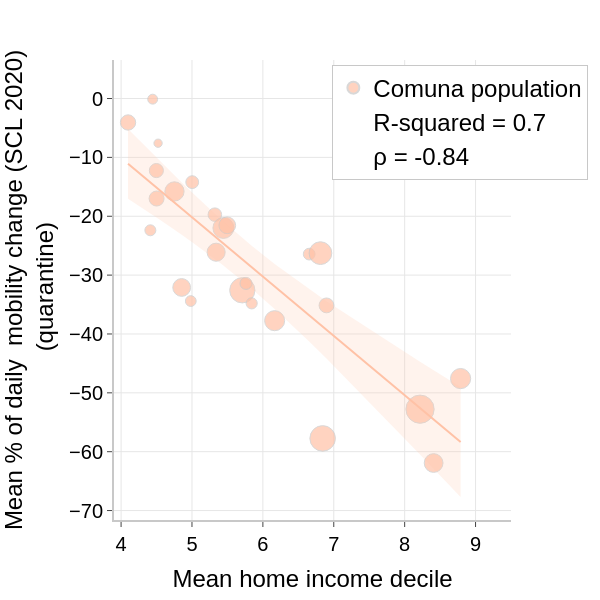} 
        \caption{IM change 2020 (for quarantine periods)} \label{fig:IM_re_quarantine}
    \end{subfigure}
    \begin{subfigure}[t]{0.24\textwidth}
        \centering
        \includegraphics[width=\linewidth]{figures/daily_IM_re_2017.png} 
        \caption{IM change 2017} \label{fig:IM_re_17}
    \end{subfigure}
    \hfill
    \begin{subfigure}[t]{0.24\textwidth}
        \centering
        \includegraphics[width=\linewidth]{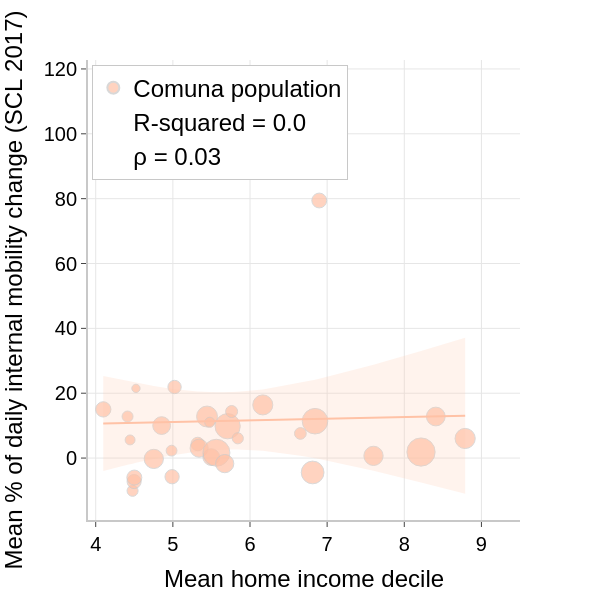} 
        \caption{IM internal change 2017} \label{fig:IM_internal_change_17}
    \end{subfigure}
    \hfill
    \begin{subfigure}[t]{0.24\textwidth}
        \centering
        \includegraphics[width=\linewidth]{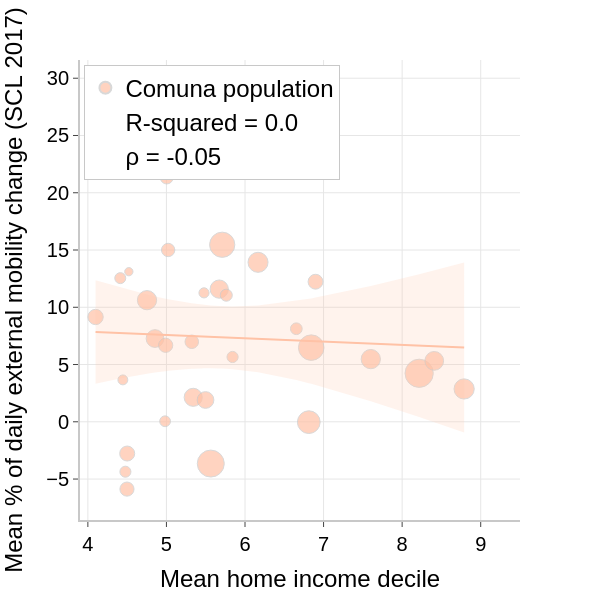} 
        \caption{IM external change 2017} \label{fig:IM_external_change_17}
    \end{subfigure}
    \hfill
    \begin{subfigure}[t]{0.24\textwidth}
        \centering
        \includegraphics[width=\linewidth]{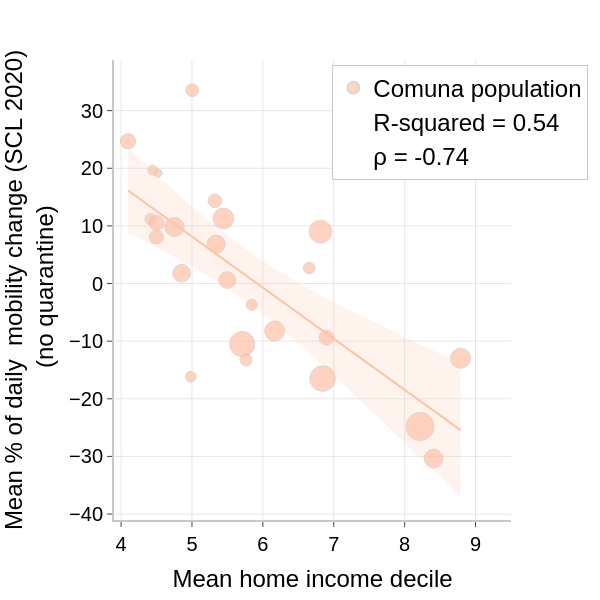} 
        \caption{IM change 2020 (for non quarantine periods)} \label{fig:IM_re_no_quarantine}
    \end{subfigure}
    \caption{Daily Mobility Index for comunas in SCL}
    \label{fig:daily_IM_2017_2020} 
\end{figure*}

\paragraph{In 2020, there was, on average, a rise in daily mobility among the lower income brackets when compared to the higher deciles.} In a study of mobility patterns across different economic deciles, our findings underscore a dynamic shift between the years 2017 and 2020. For the year 2017, the Pearson correlation coefficient between mobility (measured as percentage points of mean reduction) and economic decile (with higher values indicating greater affluence) was a marginal -0.023, with a coefficient of determination ($R^2$) of only 0.0005 (Figure \ref{fig:daily_IM_change_2017}). This suggests that the linear relationship between mobility and economic decile was notably weak, with economic decile explaining less than 0.1\% of the variability in mobility. However, the landscape dramatically changed by 2020. The Pearson correlation deepened to -0.69, revealing a more pronounced inverse relationship. Furthermore, the $R^2$ value surged to 0.48, indicating that the economic decile accounted for approximately 48\% of the variance in mobility (Figure \ref{fig:daily_IM_change_2020}). This stark contrast between the two years highlights the impact of the association between economic conditions and mobility patterns. Specifically, in an examination of mobility patterns across different economic strata, a distinct temporal evolution was observed from 2017 to 2020. In 2017, the uppermost 20\% of economic deciles accounted for nearly 19\% of the total mobility change. However, by 2020, this contribution surged to approximately 41.51\%. This indicates a pronounced increase in the concentration of mobility dynamics within the more affluent segments of the population. In a comparative analysis of mobility patterns between the richest and poorest economic segments, we employed independent two-sample t-tests to discern any significant differences in mobility reduction. For 2017, the results yielded no statistically significant difference in mobility behaviors between the 20\% richest deciles and the 80\% poorest deciles. In stark contrast, by 2020, the p-value sharply declined ($p=0.001$), indicating a prominent and statistically significant divergence in mobility patterns between these economic segments, where more affluent comunas moved much less than less affluent ones. The analyses, together with a more fine-grained study of mobility, both internal (within the areas of interest) and external (between areas of interest), can be found in Figure \ref{fig:daily_IM_2017_2020}.

\paragraph{In 2020, there was, on average, a rise in long-term relocation among the richer income brackets when compared to the lower deciles.} In our examination of migration patterns in Santiago de Chile (SCL) over the period from 2017 to 2022, we noted clear variations. In 2017, SCL saw an estimated net migration increase, with roughly 130K individuals arriving compared to about 95K departing. By 2020, this trend appeared to reverse, with the city estimated to experience a net outflow of around 35K people as emigration rose to nearly 160K, overshadowing the approximately 125K arrivals. Note that the number of people entering Santiago in 2020 is close to 2017. It is the emigration that exploded in 2020, thus causing the negative net migration. Intriguingly, by 2022, the dynamics seemed to revert to a net migration gain, with an estimated 150K individuals entering SCL, surpassing the expected 130K emigrants. In this case, although emigration remains relatively high compared to 2017, it is the significant jump in immigration that is driving the reversal in net migration. This shift hints at a renewed allure or other potential factors drawing people post-pandemic and some returning.

In general, we see a consistent change in the migration flow pattern from most comunas in Santiago in 2020 compared to 2017. The net migration rate for the individual comunas in SCL shows that there was an exodus from the city during COVID-19, with many comunas getting into negative values (see Figure \ref{fig:net_migration_map_2020}). Furthermore, by using 2017 as the baseline (Figure \ref{fig:net_migration_map_2017}), we see that, even in comunas with a positive net migration, the values were still below those from 2017.

Furthermore, in Figure \ref{fig:net_migration_per_year}, we show the net migration rate with comunas of SCL sorted from left to right by an increasing income average decile. Results show that the exodus from Santiago during 2020 was most significant for richer comunas like Las Condes and Providencia, which significantly increased their population contribution to multiple other regions.
Similar analyses for 2022 show that the net migration rate has since reverted, and it is somewhat coming back to pre-pandemic values in 2022 (Figures \ref{fig:net_migration_map_2022} and \ref{fig:net_migration_per_year}). 

The emigration trend (people leaving Santiago) was particularly pronounced in comunas with higher average household incomes, where the percentage of people leaving almost doubled in 2020 compared to 2017. Interestingly, the average socioeconomic level for 2017 showed only a weak correlation with the percentage of the population migrating from each comuna (Figure \ref{fig:em_MR_com_2017_income}). However, in 2020, the average household income decile of the comuna alone explains 57\% of the variance (see Figure \ref{fig:em_MR_com_2020_income}), hinting at a strong economic effect. This aligns with our regression analysis, which showed a higher $R^2$ value and coefficient for 2020 compared to 2017. In 2022, we observed a slight recovery after the pandemic, with income still being a much stronger predictor than it was in 2017 (Figure \ref{fig:em_MR_com_2022_income}). This is consistent with our regression analysis for 2022, which showed a significant relationship between the dependent variable and home income decile, although slightly weaker than in 2020. Notably, emigration from low-income comunas tended to increase in 2022 compared to both previous periods.

Overall, our findings suggest that income played a significant role in migration patterns in Santiago during the period from 2017 to 2022. The relationship between income and migration strengthened significantly in 2020, likely due to the economic impact of the pandemic, and remained strong in 2022, even as the situation began to recover.

\begin{figure*}[ht]
    \centering
    \begin{subfigure}[t]{0.32\textwidth}
        \centering
        \includegraphics[width=\linewidth]{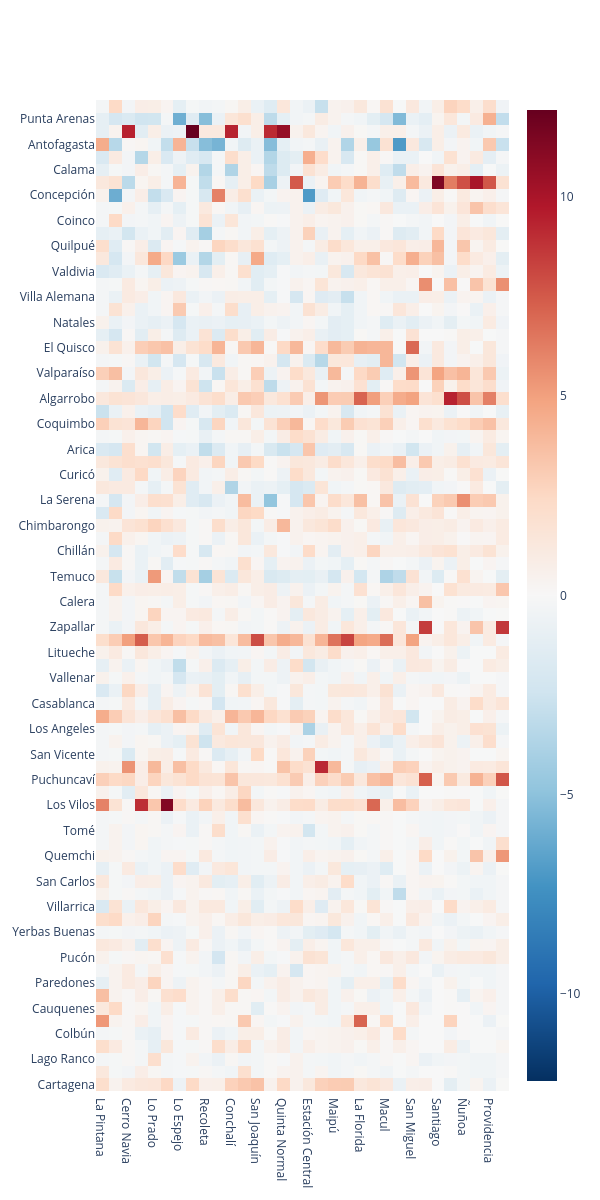} 
        \caption{Difference in Emigration from SCL in 2020 compared to 2017} \label{fig:origin_dest_matrix_diff_2020}
    \end{subfigure}
    \hfill
     \begin{minipage}[b]{.32\textwidth}
       \centering
       \vspace*{\fill}
        \begin{subfigure}[t]{\textwidth}
            \centering
            \includegraphics[width=\linewidth]{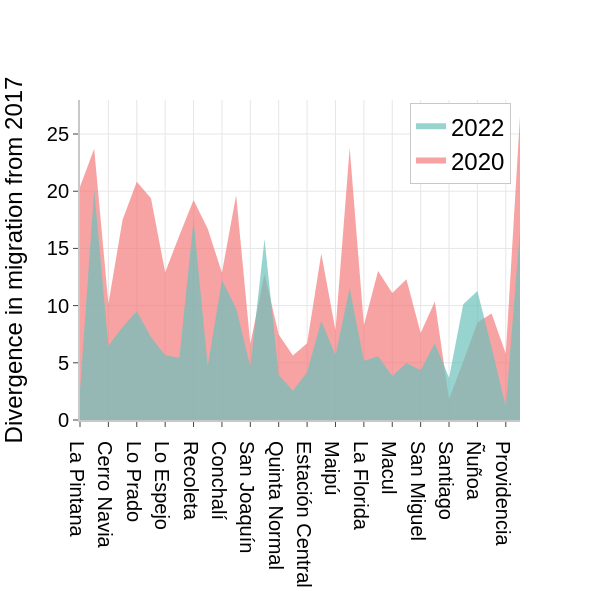} 
            \caption{Divergence in the destination preferences compared to 2017} \label{fig:destination_divergence}
        \end{subfigure}
        
        \begin{subfigure}[t]{\textwidth}
            \centering
            \includegraphics[width=\linewidth]{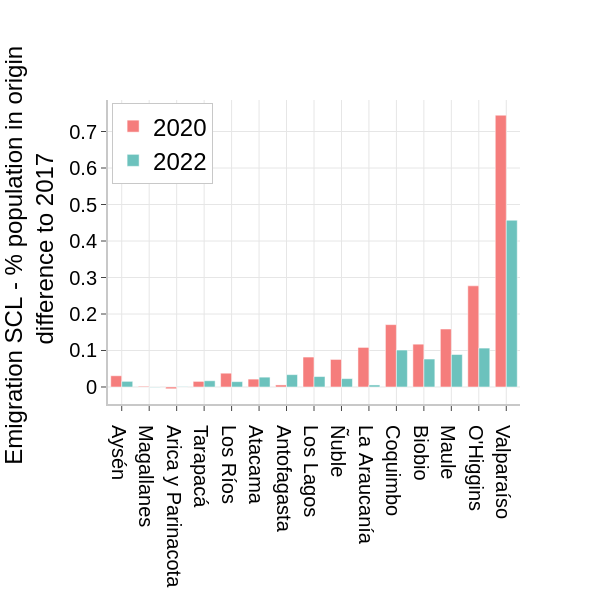} 
            \caption{Percentage of the population in the origin that represents the emigration from SCL. Difference compared to 2017.} \label{fig:percentage_origin_emigration_SCL_diff_2017}
        \end{subfigure}
   \end{minipage}
    \hfill
    \begin{subfigure}[t]{0.32\textwidth}
        \centering
        \includegraphics[width=\linewidth]{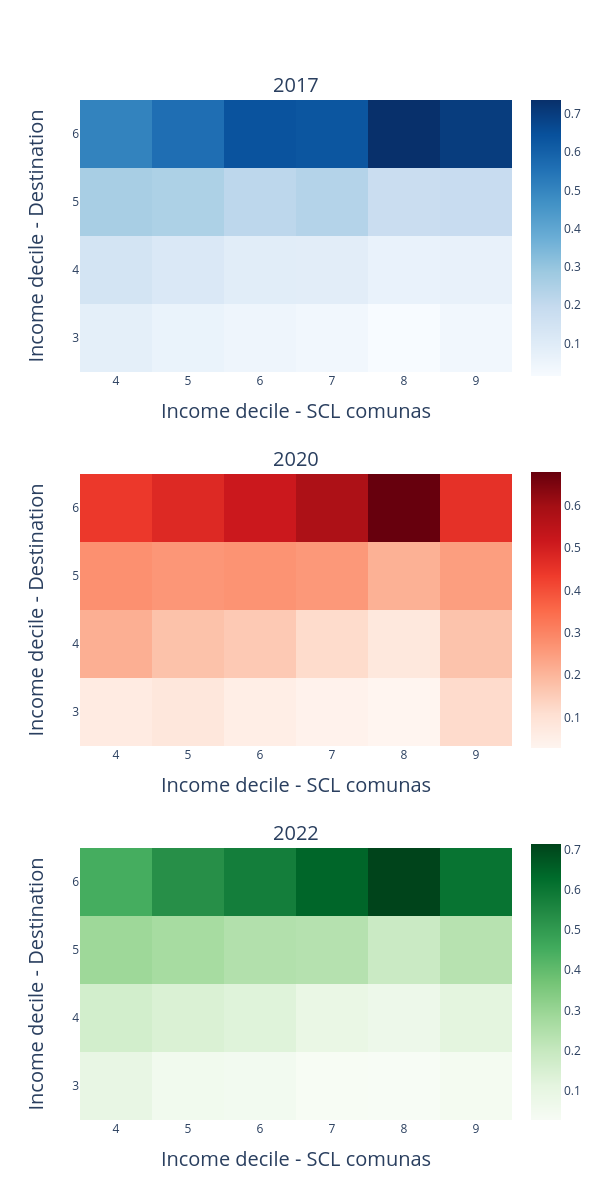} 
        \caption{Average income decile for comunas of origin and destination. Values normalized by columns.} \label{fig:decil_decil_2020}
    \end{subfigure}
    \caption{(\textbf{Destination}): Analysis of the destination for emigration from the Metropolitan Region (SCL) during the COVID-19 pandemic (2020). In (a), rows in the heatmap are sorted (from bottom to top) by ascending comunas' average household income decile. We are only including rows for which there is at least one $zscore > 1.96$. For (a) and (b), comunas from SCL in the X axis are sorted (from left to right) by ascending comunas' average household income decile.}
     \label{fig:emigration_dest}
\end{figure*}

\paragraph{In contrast to 2017, in 2020, economic aspects seem to have influenced not only the decision to emigrate but also the selection of the destinations.} We analyzed the difference in the percentage of migration at the comuna-comuna level comparing the years 2017 and 2020 for people leaving Santiago (see Figure \ref{fig:origin_dest_matrix_diff_2020}). The origin-destination matrix, sorted from bottom to top by ascending average income decile of the destination, and left to right by average income decile of the origin, shows the effect of the economic dimension on the availability of destinations for long-term relocation. People emigrating from lower-income comunas in Santiago seem to have distributed over a reduced number of destinations and tended to avoid the more expensive comunas. This is shown in the graph by a predominance of blue cells as we move to the top-left corner of the matrix (which signifies that fewer people moved between these comunas in 2020 compared to 2017). As we move to the higher-income origins (right in the matrix), we see a relative upsurge in migration over an increasing range of destinations (signaled by a predominance of red tones for the entire columns).

Utilizing the Index of Quality of Life in Urban Areas (ICVU) \cite{icvu_2020}, we evaluated the willingness among high-income migrants to cede urban amenities (as measured by the ICVU) when leaving Santiago. Given that Santiago de Chile ranks among Chile's most urbanized areas, it is noteworthy that most destination comunas registered lower ICVU scores in comparison to affluent comunas in SCL. Our analysis revealed heterogeneity in amenity trade-offs. For instance, Providencia (ranked second in ICVU) exhibited a statistically significant deviation from the expected amenity trade-offs with an average ICVU difference greater than predicted. Meanwhile, Vitacura (top-ranked) displayed less variation and stayed below the expected difference (i.e., migrants conceded less in quality of life).

Further quantification of these tendencies can be observed in the origin-destination matrices, segmented by average income decile (Figure~\ref{fig:decil_decil_2020}). Similar to the patterns observed in 2017, migration predominantly occurred towards comunas with elevated socioeconomic indices. However, in 2020, the data indicated an increased variation in amenity trade-offs across income deciles. For example, in 2020, origins with lower income deciles had a greater percentage of people moving to destinations with a closer average economic level.

To further measure the changes in the selection of destinations compared to pre-pandemic patterns, we calculate the \textit{1-Wasserstein} divergence \cite{panaretos2019statistical} for each column in the origin-destination matrices of 2020 and 2022 against 2017 (see Figure \ref{fig:destination_divergence}). We see that the divergence from 2017 is larger for comunas with lower income. Despite having the most significant increase in emigration, high-income comunas had the lowest variation in their selection of destinations. As observed before, 2022 shows again a move back to pre-pandemic patterns.

At the regional level, preferred destinations as per the hosted proportion of migrants from SCL stayed relatively stable, except for Valparaíso (a neighboring region and a popular vacation destination), which saw a statistically significant increase in immigration (see Figure \ref{fig:percentage_origin_emigration_SCL_diff_2017}). This pattern coincides with findings in other cities (e.g., New York City), suggesting that many urban residents moved to neighboring areas, second residences, and holiday destinations \cite{quealy2020richest}. Moreover, in Figure \ref{fig:percentage_origin_emigration_SCL_diff_2017}, we sorted the regions using the Gravity Model (i.e., the population at the destination over distance to SCL) \cite{poot2016gravity}. The graph shows that higher values from the Gravity Model correlate to higher percentages of migration to the corresponding region. This indicates that, besides economic aspects, distance and population density may have been still meaningful determinants in the selection of the destination in 2020 \cite{stillwell2010interaction}.

\paragraph{Despite the increase in emigration from Santiago during the pandemic, there is no evidence of preference for rural over urban destinations.}
The COVID-19 pandemic spurred distinctive migration trends globally. While some large metropolitan areas saw a surge in residents relocating to rural settings \cite{gonzalez2022rural,borsellino2022regional}, anecdotal evidence also highlighted urban relocations \cite{marsh2020escape}. In Santiago de Chile (SCL), our analysis suggests that, contrary to some trends, people predominantly favored urban comunas over rural ones.

\begin{figure*}[ht]
    \centering
    \begin{subfigure}[t]{0.32\textwidth}
        \centering
        \includegraphics[width=\linewidth]{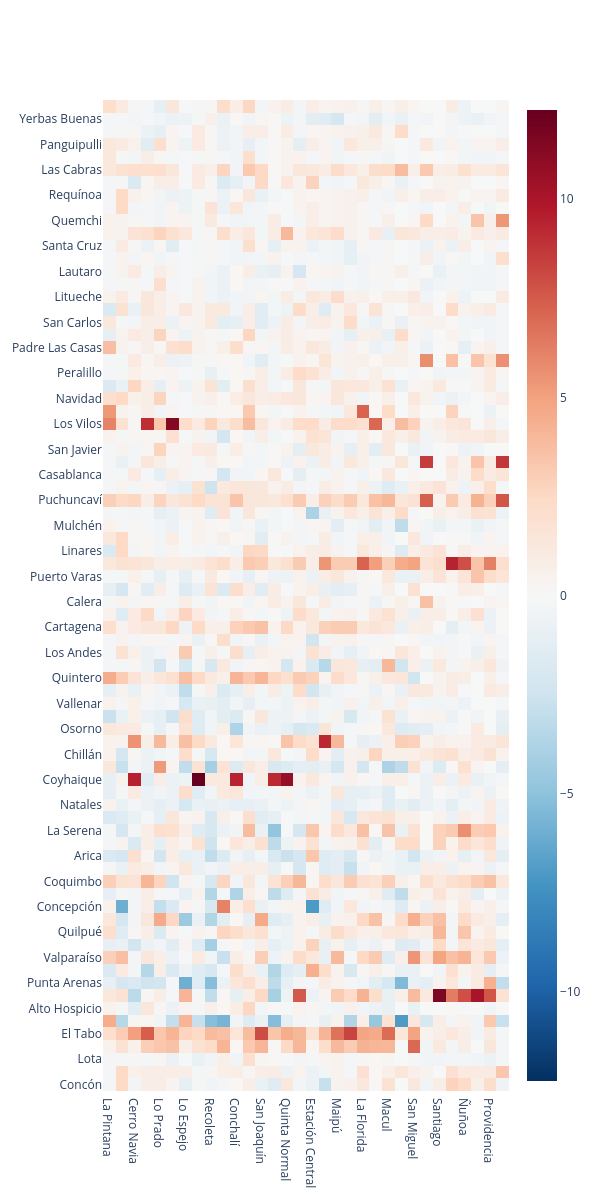} 
        \caption{Difference in Emigration from SCL in 2020 compared to 2017} \label{fig:em_com_com_2020-2017_rurality}
    \end{subfigure}
    \hfill
    \begin{minipage}[b]{.32\textwidth}
       \centering
       \vspace*{\fill}
        \begin{subfigure}[t]{\textwidth}
            \centering
            \includegraphics[width=\linewidth]{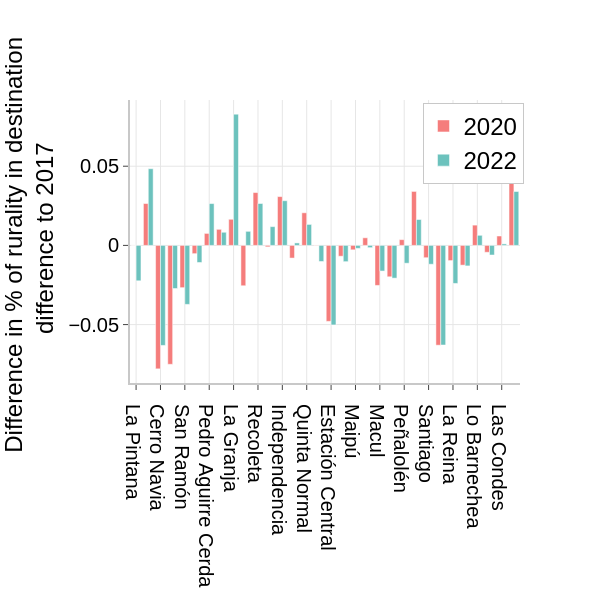} 
            \caption{difference in percentage of rurality of destination compared to 2017} \label{fig:em_MR_com_years_dest_rurality}
        \end{subfigure}
        
        \begin{subfigure}[t]{\textwidth}
            \centering
            \includegraphics[width=\linewidth]{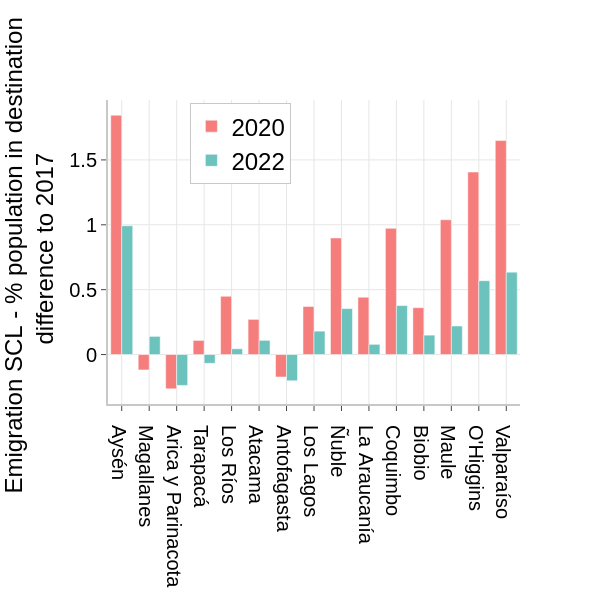} 
            \caption{Percentage of the population in the destination region that represents the emigration from SCL. Difference compared to 2017.} \label{fig:percentage_destination_emigration_SCL_diff_2017}
        \end{subfigure}
   \end{minipage}
    \hfill
    \begin{subfigure}[t]{0.32\textwidth}
        \centering
        \includegraphics[width=\linewidth]{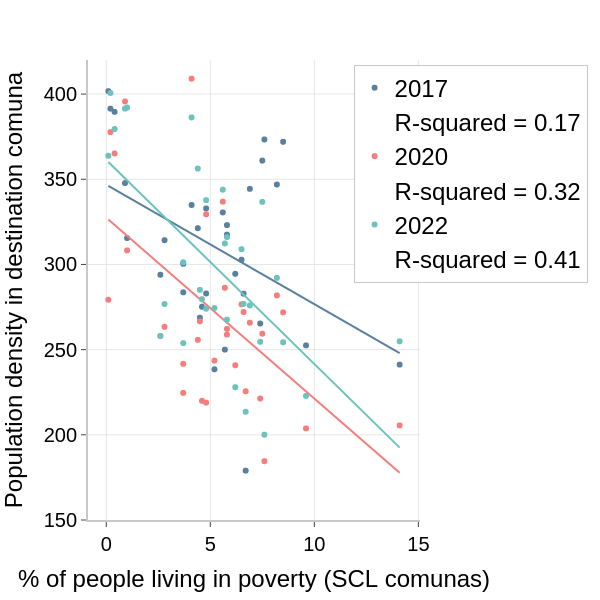} 
        \caption{Relation between emigration from SCL and the weighted average population density in the comunas of destination.} \label{fig:pobreza_pop_density_years}
    \end{subfigure}
    \caption{(\textbf{Rurality}): Difference in estimated emigration from the Metropolitan Region at the comuna-comuna level compared to 2017 according to the Internal Migration Mobile Model. In (a), rows in the heatmap are sorted (from bottom to top) by ascending comunas' percentage of the rural population. We are only including rows for which there is at least one $zscore > 1.96$. In (b), the percentage of rurality is weighted by the percentage of emigration to each destination. In both figures, comunas from SCL in the X axis are sorted (from left to right) by ascending comunas' average household income decile.}
     \label{fig:emigration_rurality}
\end{figure*}

To quantify this preference, we analyzed the origin-destination migration matrices from 2020 and 2017, focusing on the difference in migration percentages at the comuna level. Sorting the rows (destinations) by increasing rurality percentages (Figure \ref{fig:em_com_com_2020-2017_rurality}) and columns by average income decile of SCL comunas, the visualization reveals a pronounced decrease in 2020 migrations to high urbanization comunas compared to 2017. However, there wasn't a marked inclination for rural destinations. Notably, the wealthier comunas (as seen in the rightmost columns) showed increased migrations, irrespective of the destination's rurality.

To gain additional quantitative insight into the change in preference for rural destinations, we calculated the difference against 2017 in average rurality of migration destinations for each SCL comuna $\Delta R_x$, weighted by the percentage of emigrants (see more details in the Methodology Section).

The rurality preferences across Santiago de Chile (SCL) comunas are quantitatively depicted in Figure \ref{fig:em_MR_com_years_dest_rurality}. Notably, high-income comunas such as Vitacura exhibit a positive change in average destination rurality \(\Delta R_x > 0\), contrasting with over half of the comunas that either retained a similar preference \(\Delta R_x \approx 0\) or skewed towards more urbanized destinations \(\Delta R_x < 0\). These findings were observed while sorting the SCL comunas by ascending average income decile. Despite this binary urban-rural classification, no statistically significant correlation was evident between the socioeconomic status of the origin comuna and the shift in average destination rurality.

To refine this analysis, we transitioned from a binary rurality measure to a continuous metric based on population density. Figure \ref{fig:pobreza_pop_density_years} delineates the relationship between the poverty rate in SCL comunas and the average population density of destination comunas. A stronger trend emerged in 2020, indicating migration from wealthier comunas to more densely populated areas, substantiated by an increase in the coefficient of determination \(R^2\) from 0.17 in 2017 to 0.32 in 2020. Intriguingly, 2022 data revealed a further intensification of this pattern, as evidenced by an \(R^2\) value of 0.41, deviating from other metrics that reverted to pre-pandemic levels.

Finally, at the regional level, we analyzed the potential impact for the hosting regions of this increased emigration from SCL compared to 2017. We quantified the difference in the percentage of population change for each destination region and sorted them using the Gravity Model (Figure \ref{fig:percentage_destination_emigration_SCL_diff_2017}). We see that for neighboring regions such as Valparaiso and O'Higgins, despite already having relatively large populations, the emigration from SCL in 2020 still represents an increase of around 1.5\% of their normal inflow of 1.5\% - 2.0\% seen in 2017. In other words, these regions saw an almost two-fold increase in their population growth compared to previous years. Most notably, the region of Ays\'en, in the south of Chile, witnessed the biggest increase with respect to its population, going from a new 1.3\% population formed by migrants from SCL in 2017 to 3.2\% in 2020.

\section{Discussion}
Our findings illuminate distinct patterns in human mobility during the pandemic, characterized by a significant divergence between daily and long-term mobility behavior, especially within different socioeconomic strata. The dynamics underlying these patterns warrant further investigation, particularly to understand how these changes influence social, economic, and health-related outcomes in communities. 

Interestingly, the relationship between socioeconomic level and migration showed a substantial shift from 2017 to 2020. While the average socioeconomic level for 2017 had only a weak correlation with the percentage of the population migrating from each comuna, in 2020, the average household income decile of the comuna alone accounted for 57\% of the variance. This shift points to a potential change in the factors influencing migration decisions and suggests that the economic impact of the pandemic could play a role. Future research should further examine the drivers of this change and their implications for migration policy and practice.

Despite witnessing the most substantial increase in emigration, high-income comunas exhibited the least variation in their selection of destinations.
This finding implies a lower impact on their choice capacity due to the pandemic.
Low-income comunas might have been more restricted in their selection, also hinting at a potential shift in the motivation to migrate.
The factors contributing to the divergence in destination selection among this group could include economic opportunities, social networks, or amenities present in certain locations. 

Lastly, our analysis revealed that, during the pandemic, people generally avoided densely populated areas. However, they did not show a marked preference for rural areas. The avoidance of densely populated areas might reflect concerns about virus transmission risks. In contrast, the lack of preference for rural areas might be linked to factors such as access to amenities, services, or job opportunities. Future studies could explore this balance between health concerns and economic or lifestyle considerations in destination choice.

In the future, it will be crucial to continue monitoring these trends as unforeseen circumstances like wars, natural disasters, and similar crises happen and to conduct further research on the socioeconomic and policy implications of these mobility shifts. In particular, understanding the drivers of these migration patterns can inform interventions to address the needs of different communities and mitigate the potential adverse effects of such population movements. Additionally, these findings underscore the value of mobile phone data in studying human mobility, suggesting that such data could be leveraged in future research to gain further insights into migration trends and their impacts on society.

\section*{Acknowledgment}
E.E. received funding from the European Union’s Horizon 2020 research and innovation programme under grant agreement No 101021866 (CRiTERIA). L.F. and L.B. thank the funding and support of Telefónica R\&D Chile and CISCO Chile. This research was supported by FONDECYT Grant N°1130902 to L.B. and FONDECYT Grant N°1221315 to L.F.

\printbibliography

\end{document}